\documentclass{iaus}
\usepackage{graphicx}
\usepackage{natbib}

\def\chandra{{\sc Chandra}}
\def\xmm{{\sc XMM-Newton}}
\def\swift{{\sc Swift}}

\title[Workshop: X-ray Time Domain Surveys] %% give here short title %%
{The Future of X-ray Time Domain Surveys}

\author[D. Haggard \& G.R. Sivakoff]   %% give here short author list %%
{Daryl Haggard$^1$ \and Gregory R. Sivakoff$^2$}

\affiliation{$^1$Center for Interdisciplinary Exploration and Research in Astrophysics (CIERA), \\ Department of Physics and Astronomy, Northwestern University, Evanston, IL 
USA \\ email: {\tt dhaggard@northwestern.edu} \\[\affilskip]
$^2$Department of Physics, University of Alberta, %\\ 
Edmonton, AB, Canada \\email: {\tt sivakoff@ualberta.ca}}

\pubyear{2011}
\volume{285}  %% insert here IAU Symposium No.
\pagerange{1--8}
\jname{New Horizons in Time Domain Astronomy}
\editors{R.E.M. Griffin, R.J. Hanisch \& R. Seaman, eds.}
\begin{document}

\maketitle

\begin{abstract}
Modern X-ray observatories yield unique insight into the astrophysical time
domain.  Each X-ray photon can be assigned an arrival time, an energy and a sky
position, yielding sensitive, energy-dependent light curves and enabling
time-resolved spectra down to millisecond time-scales. Combining those with
multiple views of the same patch of sky (e.g., in the \chandra\ and \xmm\ deep
fields) so as to extend variability studies over longer baselines, the spectral
timing capacity of X-ray observatories then stretch over 10 orders of magnitude
at spatial resolutions of arcseconds, and 13 orders of magnitude at spatial
resolutions of a degree. A wealth of high-energy time-domain data already
exists, and indicates variability on timescales ranging from microseconds to
years in a wide variety of objects, including numerous classes of AGN,
high-energy phenomena at the Galactic centre, Galactic and extra-Galactic X-ray
binaries, supernovae, gamma-ray bursts, stellar flares, tidal disruption
flares, and as-yet unknown X-ray variables. This workshop explored the
potential of strategic X-ray surveys to probe a broad range of astrophysical
sources and phenomena.  Here we present the highlights, with an emphasis on the
science topics and mission designs that will drive future discovery in the
X-ray time domain.

\keywords{
accretion, accretion disks, 
stars: flare, 
(stars:) novae, cataclysmic variables, 
(stars:) supernovae: general, 
stars: winds, outflows, 
galaxies: active, 
galaxies: nuclei,
X-rays: general}
%% add here a maximum of 10 keywords, to be taken form the file <Keywords.txt>
\end{abstract}

%-------- section 1 --------
\firstsection
\section{X-ray Astronomy's Broad Reach}

X-ray data span an enormous dynamic range within astrophysical r\'egimes.  In
the coming decades X-ray observatories, in concert with instruments across the
electromagnetic spectrum, will systematically tackle the exciting ``time
domain''. They will have enough power to reveal the progenitors to gamma-ray
burst (GRBs), to probe the physics behind supernova (SN) shock breakout, to
identify and characterize tidal disruption events, to constrain models of the
accretion physics in X-ray binaries (XRBs) and active galactic nuclei (AGN),
and to determine the rates and driving mechanisms behind stellar flares and
their impact on space weather---to name a few examples. X-ray detectors are
unique in that every photon is time tagged, energy tagged, and assigned an
accurate sky position. X-ray observations also cover time-scales from
sub-millisecond to $\sim$40--50 years, span orders of magnitude in spatial
resolution, and achieve a decade in energy coverage with decent energy
resolution. The result is sensitive, energy-dependent light curves and
time-resolved spectroscopy for every target.  A wealth of high-energy
time-domain data already exist, from which variability on time-scales ranging
from microseconds to years has already been identified in a wide variety of
objects. 

In this workshop we discussed the missions that would be optimal for
discovering and characterizing X-ray transients and variables.  We were
motivated in part by the enormous interest that has been expressed by the
astronomical community (and evidenced by this very Symposium) in optical and
(more recently) radio transients. We felt that the X-ray community is not, at
present, making a compelling case for the power of X-ray observatories which
are optimized for time domain studies (with a few notable exceptions such as
RXTE and {\sc Swift}). We hope that the guiding questions outlined below,
together with those generated by workshop attendees, will bring into focus the
kinds of efforts needed to lobby most effectively for those missions, archives,
cadences and science objectives in order to ensure that X-ray astronomy is well
resourced in the future and thus able to contribute substantially to the
exploration of transient phenomena.

The time domain is already expanding rapidly.  To optimize these many transient
domain studies we must connect the targets and the science at multiple
wavelengths.  Competition can be inimical to progress; although some might fear
lest the optical and radio communities absorb resources away from the X-ray
transient community, there was general consensus that the deepest insights into
physics, and hence the highest science impact, result from coordinated,
multiwavelength observations.

This overview of the workshop does \emph{not} explore the full scope of science
accessible in the X-ray domain, nor advocate any particular mission. Both the
science and the technology are rapidly evolving, and attempts to place the
entirety of X-ray astronomy under a single umbrella may be a questionable
exercise---as explained in Martin Elvis' response to NASA's recent call for
``Concepts for the Next NASA X-ray Astronomy Mission'')\footnote{NASA has
recently solicited the community to suggest new X-ray mission concepts for
advancing the goals of the Physics of the Cosmos (POC) programme (NASA RFI
NNH11ZDA018L). These submissions are public, available on the POC webpage: {\tt
http://pcos.gsfc.nasa.gov/studies/x-ray-mission-rfis.php}.}.  Instead, we hope
to prompt the astronomical community into thinking about the central role which
X-rays have played and still can and should play, in our exploration of astronomy's
time domain.

\section{Guiding Questions}

We asked workshop attendees to discuss these guiding questions:

\begin{enumerate}
\item 
In recent years optical and radio transient science have increasingly gained
attention among the general astronomical community. At the same time, X-ray
transient surveys seem to be ceding ground, both financially and
scientifically. What are the most compelling science cases for current and
future X-ray transient studies? What efforts does the X-ray transient community
need to undertake to lobby most effectively for the importance of X-ray
transient studies (past and present) to the general astronomical community?

\item 
The Rossi X-ray Timing Explorer (RXTE) has been a tremendous boon for studies 
of X-ray transients.  However, it will cease operation at the end of this
year. While some of its scientific capacities can be shifted to current
instruments like {\sc swift} and MAXI, other capacities are unique to RXTE
among currently flown instruments.  What steps do we need to take to transition
from the era of RXTE to the era without it?  What important lessons have we
learned from RXTE? How will new planned or soon-to-be-launched instruments
support X-ray transient surveys? What inventive ways can we develop to utilize
new instruments that may not have been designed originally for X-ray transient
studies?

\item 
The scientific output of X-ray transient surveys can be greatly increased
through multi-wavelength observations. How do we best coordinate
multi-wavelength observations, especially for X-ray transient surveys? Do we
need to develop an X-ray Transient Network, or are existing infrastructures
like the Gamma-ray Circular Network and the Astronomers Telegram sufficient?
What cadences are needed to achieve various science priorities at different
wavelengths? Are there opportunities for ``citizen science'' with X-ray
transient surveys?
\end{enumerate}

\section{Workshop Highlights}

The workshop was structured as a pure discussion---there were no formal science
talks.  Some of the most active discussions that took place are outlined
below. A 1.5-hour audio recording of the workshop, together with a
written transcript, are available at \\ {\tt
http://faculty.wcas.northwestern.edu/$\sim$dha724/xray\_transients\_2011/}.

\subsection{X-ray Transients and Variables}

Our first discussion was of transients (unanticipated [dis]appearance or
flaring) as opposed to variables (periodic or repeated fluctuations). Are X-ray
studies more likely to uncover ``variables'' than ``transients''?  The majority
opinion was that most X-ray variables were initially identified as transients
(as is indeed the case with optical/radio transients), and that in most cases
the distinction is driven by the detection limits of individual surveys. For
example, on very deep optical data (to $\sim$28$^{\rm th}$ magnitude) one may
begin to see progenitors of Type Ia supernovae (which are themselves are
probably variable) in addition to novae, X-ray binaries and the like.

\subsection{The Science Case(s) for X-ray}

It is essential to state the most compelling science cases for current and
future X-ray transient studies---to identify what is unique about the X-ray
domain and why it should be compelling to fund an X-ray mission rather than a
UV or IR one.  \emph{Strong gravity} and \emph{accretion physics} are both
areas to which the X-ray time domain brings a unique view.  The most
interesting individual science cases for X-ray time-domain studies included:

\begin{itemize}
\vskip6pt
\item Gamma-ray bursts (black-hole birth, cosmological probes)
\item Supernova shock break-out
\item Tidal disruption events
\item X-ray variability of AGN and XRBs 
\item Giant hard X-ray flares (from flare stars and blazars)
\item Impact of stellar flares on space weather/planetary habitability
\item Variability in SgrA*
\item Accreting millisecond pulsars
\item Coherent pulsations and QPOs in neutron stars 
\item Galactic black-hole and neutron star populations
\end{itemize}

\vskip6pt

In addition, other X-ray variables, not yet recognized as such, might supply
the most compelling physical insights, though it is difficult if not impossible
to base an X-ray mission on only an anticipated benefit. Many of the phenomena
cited were originally discovered in the X-ray domain (though most remain only
poorly characterized).  However it was felt that, in coming years, the impetus
will most likely come not from X-ray missions but from optical or radio
telescopes, reflecting an enthusiasm for the ``new'' (LOFAR, ATA, PTF and
potentially LSST) as opposed to the ``old'' or established (RXTE All-Sky
Monitor, {\sc swift}, \xmm and {\sc Chandra} surveys). If we make out that the
X-ray sky is a known entity, then the potential for discovery is perceived to
be greater at less known wavelengths, making the latter seem more exciting.  An
X-ray transient mission therefore needs some goal like testing general
relatively to bolster its case, i.e., something that can only be done through
X-ray science.

X-ray variability in AGN and XRBs probes the physics of the
inner accretion disk. These, in particular, test strong gravity. The same is
true for tidal disruption events. The structure of the variability and its
time scale may assist in distinguishing between radiatively efficient and
inefficient accretion flows and the mechanisms responsible for launching jets
and winds. Sensitivity to very rapid variations (coherent pulsations, QPOs,
fractional variability) is critical for understanding local XRB sources, and
may shed light on more distant sources by analogy.

From the multiwavelength perspective, radio quenching and radio flaring have
been seen in X-ray binaries within days. Hence, having missions that have the
capacity to observe an XRB daily after an outburst has proved 
critical. The difficulty now is coordinating efficiently with other
observatories; in the radio (for example) coordinations with EVLA have 
improved with the introduction of dynamic scheduling, but are still of the 
order of a few hours.

For AGN the relevant time scale is weeks to years.  As pointed out, AGN go
into a deep low state and stay there for days or weeks; that is when the X-ray
spectral complexity is most pronounced, and when distinguishing between
different inner disk models is most effective.  Thus, for AGN it might be the
dips in their light curves, not the flares, which prove more interesting.
Monitoring tidal disruption flares is also most effective on time-scales of
weeks, but time-scales of minutes have not yet been explored for blazars. At
GeV energies we are limited by statistics, but there is sub-day variability,
and presumably it is the X-ray non-thermal component that is varying.
As described in a contribution by S. Kulkarni,
%by \pageref{Kulkarni}, 
the time-scale for the X-ray shock break-out from supernovae is hours or less.

\begin{figure}[b]
\vspace{-0.7 cm}
%\vspace{-0.5 cm}
\begin{center}
 \includegraphics[width=5in, angle=0]{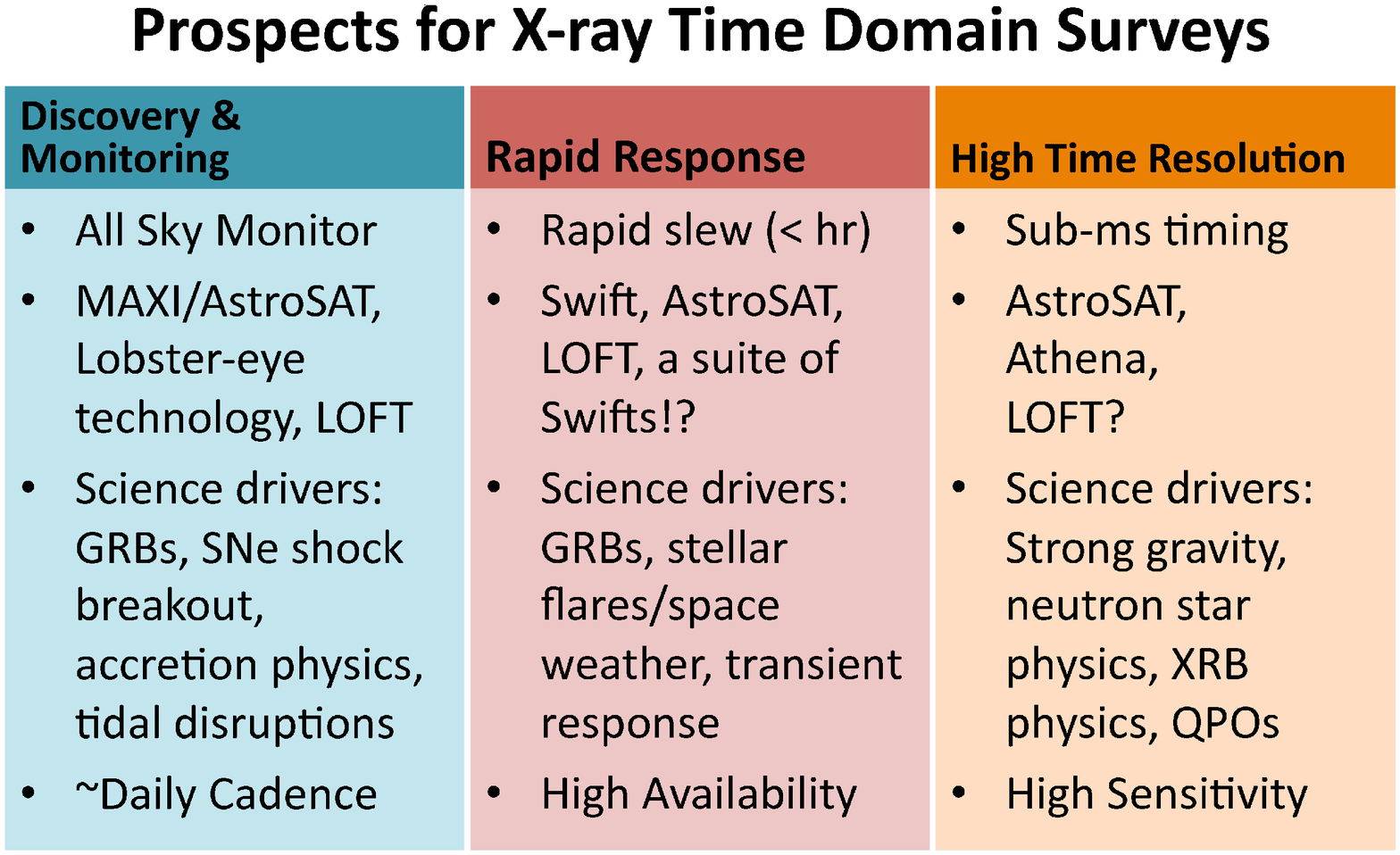} 
% \vspace*{-1.0 cm}
% \caption{}
   \label{fig1}
\end{center}
\vspace{-1.3 cm}
\end{figure}

In stellar coronal variability, both sensitivity and wavelength coverage are
important. Greater sensitivity
allows one to look for flares from stars at greater distances or for weaker
flares from stars less distant, but we require the multi-wavelength context
(soft X-rays, hard X-rays, UV/optical) to facilitate a full
interpretation. X-ray emission from stars provides information about the
coronal material and the coronal dynamics, and which cannot be obtained from
other wavelengths; X-rays show how the tenuous coronal plasma is reacting to
magnetic reconnection.  There are many aspects regarding stellar flares that
are of outstanding interest.  What drives the extreme energy release?  How do
flares affect the stellar environment (both in the context of young stars where
planets are forming in a disk, and for older stars where planets have already
formed)?  How might flares affect habitability?  Flares need to be understood
in the context of larger magnetic processes, and dynamo processes. To study
stellar flares in detail requires high time-resolution, e.g., responses within
minutes, because most of the energy in the initial flare is released in the
first few minutes in the so-called ``impulsive'' phase, when one expects to see
hard X-rays and radio emission; later the flare transitions to the ``gradual''
phase when thermal X-rays and the UV/optical responses begin to dominate.
Observations at different time-scales thus probe different physics.  Statistics
of stellar flares can usefully be derived on all times-scales: minutes, hours
and days.

Another extension of stellar flare science involves space weather and the
impacts upon the Earth.  Studying the solar corona might in principle teach a
great deal about flares on just one class of star but would teach little about
its past and projected future behaviour; broadening the sample to many
different types of star suggests how the Sun behaved in the past, and how it
might behave in the future.  The inverse of the argument is to regard
observations of stars as proxies for modelling how the Sun's influence on space
weather might evolve.  Strength might then be given to potential new missions
by opening them to other scientific communities (and their resources), though
careful crafting of the science case would be imperative.

Clearly, the cadences required (minutes to years) depend crucially on the class
of sources being explored.  Different science goals are best accomplished with
different technologies (see Fig.~1). Very fast transients represent territory
that is largely unexplored, while at the other end of the scale all-sky X-ray
monitoring programs have mission lifetimes that are poorly matched to the long
(rest-frame) variability time-scales of AGN.  At present there is too much
reliance upon serendipity; the 1999 flare of Sgr~V4641, for instance, or the
recently reported outburst in the Arches cluster could easily have been missed.

\subsection{Multi-wavelength Coordination}

At several junctures the workshop discussed practices for coordinating
multi-wavelength observations and sending alerts to the community.  For
example, is there a need to develop an X-ray Transient Network, or are existing
infrastructures like The Gamma-ray Circular Network and the Astronomer's
Telegram (ATel) sufficient?  
%(see \pageref{VO}).  
One existing problem is a
degree of confusion in nomenclature.  If different groups use different names
to identify the same sources, it results in complications and leads to
duplicate follow-up observations.  This seems to be particularly true when the
Galactic Centre is up and is being observed by {\sc Integral}.  A ``transient
wiki'' could keep track of everything that is currently active.  An increase in
the number of joint proposals allowed (e.g., NASA$+$ESO or
ESA$+$NOAO/Australian facilities) might also be important.

In general, ATels and other alert services seem to be serving the community
well. Moreover, inside ATel there is now the AtelStream, which is a scheme for
unifying announcements.  There is also no doubt that the situation regarding
joint proposals has improved tremendously over the past decade, but the need
for a continued push for time-share agreements and joint proposal
opportunities, especially for projects which require strictly simultaneous
data, is strongly supported. In practice, it currently requires a significant
commitment of time to coordinate a multi-wavelength campaign, possibly because
of identifyable structural issues: very few observatories are set up for
multi-wavelength collaborations.  One successful example is the excellent
inter-agreement between SAO/{\sc Chandra} and NRAO/EVLA, within which it is
quite straightforward to obtain simultaneous X-ray and radio observations.  Two
modes are involved: the ``discovery mode'' for the transients, requiring rapid
slew and other time-critical follow-ups, and the ``follow-up'' mode when
multiple instruments need to bear down on the same target.  The latter mode
requires either robotic streams or actual structural changes to the way in
which time is granted and/or scheduled.

Multi-wavelength follow-up of X-ray targets can also suffer from a mismatch in
timing resolution.  For example, X-ray data are time-tagged, and events can be
resolved easily at the millisecond level, but that information is of little
help when trying to coordinate those data with an IR observation, where
integrations run for minutes or longer, and the outcome is a comparison of two
completely different time domains.  One solution might be to use large-format
optical/IR photon-counting detectors which automatically incorporate time
tagging, discussed in contributions from K. O'Brien and B. Welsh.
%by \pageref{O'Brien} and \pageref{Welsh}. 
Absolute timing stamps can also be incorporated.  However, very high 
time-resolution detectors generate enormous quantities of data and huge files.

\subsection{Optimizing Existing and Future X-ray Missions for Time Domain 
Science}

The workshop discussed the following past, present, and future X-ray missions
in detail and how they might accomplish the science goals outlined above:

\vspace{0.3cm} 
\noindent{{\bf Planned/Proposed}:
	NuSTAR,
	AstroSAT,
	ASTRO-H, 
	eROSITA,
	GEMS, %Gravity and Extreme Magnetism SMEX
	SVOM, \linebreak %Space-based multi-band astronomical Variable Objects Monitor
	Athena,
	LOFT, %Large Observatory For X-ray Timing
	WFXT, %Wide Field X-ray Telescope
	JANUS, %Joint Astrophysics Nascent Universe Satellite
	Lobster, 
	Smart-X, also earlier footnote
	%EXIST
	%IXO
	%Gen-X
	%Pharos
	}\\
\noindent{{\bf Active ($+$Archival)}:
	\chandra,
	\xmm,
	{\it Suzaku},
	\swift,
	INTEGRAL,
	MAXI
	}\\
\noindent{{\bf Archival}:
	RXTE, %Rossi X-Ray Timing Explorer 
	ROSAT,
	{\it Einstein}
	}
\vspace{0.3cm}\\

The recording and transcript include descriptions of individual missions; see
also contribution from N. White.
%\pageref{White}.  
The instruments which are now current are also providing
extensive archives that will be particularly useful for time-domain studies
involving longer baselines.

The relative merits of an X-ray all-sky monitor, rapid slew missions and
missions or instruments optimized for high time-resolution came in for
considerable discussion.  There had been broad support at a HEAD meeting 11
years ago for an X-ray all-sky monitor, but as the demand for sensitivity
increased the payload grew, and soon it faced much stiffer competition as a
stand-alone mission---and lost. The landscape may be different now owing to the
rapid growth of, and huge investment in, ground-based programs like the Palomar
Transit Factory (PTF) and PanSTARRS, and the Large Synoptic Survey Telescope
(LSST) promised in the next decade.  When the radio equivalent is also added,
the demand for X-ray all-sky sensitivity at least an order of magnitude better
than present values will surely increase.  Among the missions that might fill
that niche are Janus and Lobster-eye detectors, described in 
%\pageref{White}.
contribution from N. White.

In discussing the lessons learned from RXTE, the workshop recognised that
flexibility in responding to target of opportunity requests (ToO) is critical
for X-ray timing studies, that the discipline needs a capability to observe how
the timing properties themselves change in time (they are sharper probes than
changes in the energy spectra), and that some of the work done by RXTE---in
particular the all-sky monitoring---can be done in the optical and infrared
from the ground because most X-ray binaries (except the highly extincted ones)
show enhanced optical and infrared emission during outburst.  {\sc Swift} has
been fantastic in its rapid response to ToOs and can take over nicely from RXTE
in certain r\'egimes, but lacks the effective area for RXTE's timing work,
specifically for the study of pulsations and QPOs.  It is possible that the
Indian mission AstroSAT (the launch is planned in 2012) will recover many more
of RXTE's capabilities, and may even improve on them through its increased
sensitivity. The AstroSAT data will be proprietary, but the possiblity for
real-time release of transients remains open.  The data archive
will be housed at the Inter-University Centre for Astronomy and Astrophysics
(IUCAA), but the plans for access to the data are unclear.

Data access and availability of funding, particularly for serendipitous and
archival studies, influence which science and which missions gain traction in
the astronomical community. One drawback of mission designs like that of {\sc
Swift} is its lack of funding for scientists pursuing ToOs.  A similar problem
affects the many X-ray mission archives such as ROSAT,
\chandra, \xmm, etc., that could be used for transient
and variability science, as well as the utilization of multi-wavelength
archives like GALEX and SDSS.  Unfortunately, since most archival research is
funded through soft money, competition for that funding influences the type of
science that gets done since proposals need to be tailored to the preferences
of the funding agencies.  One possible funding programme is NASA's {\it
Research Opportunities in Space and Earth Sciences} programme, which funds
research connected with NASA missions, including {\sc Fermi}, \chandra and
GALEX.  Indeed, radio astronomers studying compact objects could access NASA
monies to do the radio follow up. In the EVLA's model for data sharing, the
so-called RSRO time (Resident Shared Risk Observing), an observing team could
be awarded pre-commissioning time in return for at least one team expert taking
``in residence'' status at the facility, but such a programme is unlikely to be
workable within a space-based context.  However, pipeline and software
development was proposed as one area where data exchange might be feasible.

Future advances are likely to require yet higher time resolution and higher
energies. The proposed Large Observatory For X-ray Timing (LOFT; possible
launch $\sim$2020) is a high-sensitivity time-domain mission that could be the
sort of instrument required; one possible science driver could be the spectral
timing of black holes and AGN.  An alternative might be a vast improvement in
``gamma-ray burst'' type capabilities, such as an instrument with the solid
angle of BAT but 10 times more sensitivity, better source localization, and
with an IR telescope; it would open up a huge phase space which has never been
probed before.  Such a mission would specifically support a range of science
projects, from SN shock break-outs and tidal disruptions to moderate redshift,
gamma-ray bursts at $z$ $>$ 9 and searches for the periods of ultra-luminous
X-ray sources.  Meanwhile, the Wide Field X-ray Telescope, a proposed
medium-class NASA mission, could be a powerful instrument for transient
detections in the distant universe and could consider targeting the LSST
Deep-Drilling fields repeatedly during its lifetime.

\subsection{Opportunities for Citizen Science}

As a final topic for discourse, the workshop explored possibilities to involve
citizen science in X-ray transient studies.  So far there has been little
involvement by non-specialists in X-ray or high-energy programmes.  However,
citizen science is rapidly becoming recognised as a way of getting interesting
science done and---more importantly---of engaging the public and also
achieving certain tasks that need to be carried out in order to justify the
investment in support of science.

It was felt that the amateur community who normally worked in the optical
domain would be enthusiastic about following up X-ray transients.  Help could
be enlisted through a message to the American Association of Variable Star
Observers (AAVSO) or similar organization, seeking observers willing to follow
a 14, 15 or 16$^{\rm th}$ magnitude object.  Even though it is unlikely that
the faintest targets could thus be followed, the benefit to X-ray science
would be the adaptation of abundant capabilities across multiple wavelengths. 
Indeed, many of the AAVSO data are of exceptionally good quality, and amateur
observers collectively have the advantage of wide longitudinal coverage,
something which is not possible for many professional astronomers.

It should be recognized, however, that searches which make extensive use of
existing data can be computationally intensive---for instance, if one tried to
find every possible transient in the INTEGRAL or BAT archives, or looked for
transients in the
\chandra\ and \xmm\ deep fields on every possible timescale. 

Such archival searches are often very RAM intensive and may not be adaptable to
software that runs on unused cycles in the same way that (say) SETI@home or
Einstein@home can be run.  Another suggestion was to coordinate amateurs to
monitor dense regions of the sky in some systematic way in order to observe new
X-ray binaries in outburst.  Many of those systems rise quickly to
$\sim$16$^{\rm th}$ magnitude, and nowadays that is within reach for a large
number of amateurs.

\section{Summary}

The X-ray time domain uniquely probes strong gravity, accretion physics,
supernova shock break-out and stellar flares. Specific tests of the first two
involve the inner accretion disks of X-ray binaries and AGN. Changes in the
X-ray variability and in its time-scale probe the structure of accreting
degenerate systems, and increased sensitivity to rapid variations enable
studies of XRB pulsations, QPOs, and fractional variability.  The spectral
timing of black holes and AGN may also reveal the structure, and deep low
states in AGN may give a particularly clean glimpse into their spectral
complexity. Shock break-out has been an exciting topic that has featured
throughout the Symposium, and the race is on for the first observations of an
X-ray shock break-out.  X-ray emission from stellar flares probe coronal
material and its dynamics.  Most of the energy from the initial flare appears
in the X-ray and radio, usually within a few short minutes. Stellar and solar
flares are critical to our understanding of space weather, and may have a
profound impact on the habitability of planets.

\vspace{0.5cm}
%\section{Acknowledgements}
\noindent{{\bf Acknowledgements}}
\vspace{0.1cm}

We thank the 30 or so scientists who attended this workshop and contributed to our discussion, and in particular we thank Phil Charles, Stephane Corbel, Boris Gaensicke, Stefanie Komassa, Shri Kulkarni, Ashish Mahabal, Roberto Mignani, Rachel Osten, Danny Steeghs, Tom Vestrand, Barry Welsh, Peter Williams, and Patrick Woudt for generating lively dialogue. We also appreciate input from Niel Brandt, Craig Heinke, 
Tom Maccarone, and Richard Mushotzky, who commented on our guiding questions though 
they were not able to attend.

\end{document}